\documentclass[twocolumn,preprintnumbers,amsmath,amssymb]{revtex4}
\usepackage{graphicx}

\usepackage{amsmath}
\usepackage{amssymb}
\usepackage{amsthm}
\usepackage{amsfonts}
\usepackage{enumerate}

\usepackage{centernot}
\usepackage{tikz}

\begin{document}

\title{Nonlinear Quantum Hall effects in Rarita-Schwinger gas}

\author{Xi Luo$^{1}$, Feng Tang$^{2,3}$, Xiangang Wan$^{2,3}$ and Yue Yu$^{4,3}$}

\affiliation {${}^1$CAS Key Laboratory of Theoretical Physics, Institute of
Theoretical Physics, Chinese Academy of Sciences, P.O. Box 2735,
Beijing 100190, China\\
${}^2$National Laboratory of Solid State Microstructures and Department of Physics, Nanjing University, Nanjing 210093, China \\
${}^3$Collaborative Innovation Center of Advanced Microstructures, Nanjing 210093, China
\\
${}^4$ Center for Field
Theory and Particle Physics and State Key Laboratory of Surface Physics, Department of Physics, Fudan University, Shanghai 200433,
China }

\date{\today}

\begin{abstract}
Emergence of higher spin relativistic fermionic materials becomes a new favorite in the study of condensed matter physics. Massive Rarita-Schwinger 3/2-spinor  was known owning very exotic properties, such as the superluminal fermionic modes and even being unstable in an external  magnetic field. Due to the superluminal modes and the non-trivial constraints on the Rarita-Schwinger gas, we exposit  anomalous properties of the Hall effects in (2+1)-dimensions which subvert the well-known quantum Hall paradigms.  First, the Hall conductance of a pure Rarita-Schwinger gas is step-like but not plateau-quantized, instead of the linear dependence on the filling factor for a pure spin-1/2 Dirac gas. In reality,  the Hall conductance of the Dirac gas is of quantized integer plateaus with the unit $\frac{e^2}h$ due to the localization away from the Landau level centers. If the general localization rule is applicable to the disordered Rarita-Schwinger gas, the Hall  plateaus are also expected to appear  but  they are nonlinearly dependent on the Landau level index. Furthermore,  there is a critical magnetic field beyond which higher Landau levels become unstable. This confines the filling factor of the system.  We also show that the non-hermitness of the effective Hamiltonian is not crucial to the nonlinearity of the quantum Hall conductance.
\end{abstract}


\maketitle

\section{Introduction}

Since quantum Hall effects were discovered\cite{iqh,fqh}, a standard paradigm of condensed matter physics has been set up. Indeed, the quantum spin Hall effects\cite{km,km1} and the quantum anomalous Hall effects \cite{hald}  were subsequently unearthed \cite{qsh,qah}. 
The realm of condensed matter physics has then expanded rapidly into the {\it topological states of matter} \cite{kz,qz,wan}. Besides the non-trivial topological aspects of single particle band theories, there are also theoretic proposals for interacting symmetry protected topological  phases\cite{cglw} and topological orders\cite{ki,lw}.

The question raised now here is: Is it possible that there are some physical systems in which the quantum Hall paradigm is overthrown or amended? More specifically, what we ask is: Are there two-dimensional quantum gases whose quantum Hall conductance is not proportional to an integer or a fractional filling factor? In quantum mechanics, we had such an experience: The eigen energy of a harmonic oscillator is linearly dependent on the main quantum number while the eigen energy of the hydrogen atom is nonlinearly dependent on this integer.   

To search for such systems, one notices that all these systems studied so far are of either spinless bosonic excitations or fermionic quasiparticles with spin$\leq\frac{1}{2}$.  The breakdown of the quantum Hall paradigm seems no hope in these systems. Fortunately,  the emergence of relativistic fermionic quasiparticles with higher spin and multiple degeneracy  in metallic and insulating materials becomes the latest fast growing area in the research of condensed matter physics \cite{ly,IF,newfe,hlf,wfdf1,wfdf2,zz,hs,wkrk,ez,tlyw,note}.  The prophesies of higher spin fermionic quasiparticles with a linear dispersion in certain materials \cite{ly,IF,newfe} give a good understructure to realize Rarita-Schwinger semimetal which was predicted if a non-local external potential or a specific interaction is introduced \cite{ly}.  It is also possible to simulate the massive Rarita-Schwinger fermions in condensed matter systems \cite{tlyw}.

Can we  anticipate interesting new physics for these higher spin fermions like quantum Hall effects nonlinearly depending on the Landau level index?  On the other hand, the discussions on the measurable physical properties of these fermionic systems, especially their topological and transport properties, are important for the sustainable developments and potential applications of the research for this new field. 

The exotic and even ambiguous properties of higher spin systems are well-known in high energy physics and quantum field theory for a long time. For example, when a massive Rarita-Schwinger fermion couples minimally to an external magnetic field, a superluminal mode will emerge and the system turns out to be non-hermitian or even unstable \cite{rs1,2drs,rs2}.  Similar besetments may also appear when a massive higher spin field experiences an external potential \cite{vz,potential,potential1,potential2}. However, the superluminal propagation of an elementary particle is believed to be forbidden. Thus, the massive charged Rarita-Schwinger fermion might  not be an elementary particle
although the superluminal mode disappears if the Rarita-Schwinger fermion couples non-minimally to the electromagnetic field \cite{weinberg}.    

In a condensed matter system, since the Fermi velocity serves as the "speed of light" which is much smaller than the speed of light in the vacuum, superluminal modes with a velocity faster than the "speed of light" are allowed. Do the superluminal modes bring us any new physics? For instance,  due to the propagation of the superluminal modes, is it possible that the quantum Hall effects of the Rarita-Schwinger fermion gas in 2+1 dimensions is different from those of the Dirac fermions in graphene \cite{gra1,gra2,gra3}?

Because gapless Rarita-Schwinger equations in (2+1)-dimensions have only a trivial solution (see Discussion in \cite{ly}.),  we study the (2+1)-dimensional massive charged Rarita-Schwinger fermion gas. Due to the constraints for the massive Rarita-Schwinger fermions and the existence of the superluminal mode, the result is non-trivial. Even for a pure Rarita-Schwinger gas,  
we find that in a strong magnetic field but not strong enough to break the stability of all Landau levels, instead of a linear dependence on the filling factor in a pure Dirac gas, the Hall conductance of the pure Rarita-Schwinger gas is step like but not plateau-quantized. That is, the Hall conductance for a given Landau level decreases nonlinearly as the filling factor increases. However, a stepped jump exists between two adjacent Landau levels. In reality, because the quantum states away from the centers of Landau level are localized by disorders, the Hall conductance of the non-interacting Dirac gas is quantized to integer times of the unit $\frac{e^2}h$ as well known \cite{gra1,gra2}. If the localization rule of a two-dimensional quantum gas is also applicable to the Rarita-Schwinger gas, the quantum Hall effects with a nonlinear dependence on the integer Landau level index are expected, i.e., the Hall plateaus are not located at the integer but proportional to a nonlinear function of the integer.

When the external magnetic field is  stronger than a critical field after that the superluminal mode can propagate in all spacelike intervals, we see that higher Landau levels become unstable. This phenomenon does not become severe monotonously as the external field increases but worst appears in a specific range of the field strength.

The nonlinear Hall effects of the Rarita-Schwinger gas is caused by the  superluminal modes and the non-trivial constraints.  The effective Hamiltonian of the superluminal modes is non-hermitian. However, we show that in the limit of a small $G\propto B/m^2$ in which the effective Hamiltonian of the system is hermitian, the  quantum Hall conductance is still nonlinear to the Landau level index. This implies that the non-hermitness of the effective Hamiltonian is not crucial for this exotic phenomenon. 

This paper is organized as follows: In Sec. \ref{III}, we briefly review the Rarita-Schwinger spinor and the minimal coupling to a constant magnetic field in 2+1 dimensions. The superluminal modes of the Rarita-Schwinger fermion are introduced. With the preparations in Sec. \ref{III}, we calculate the Hall conductances of the single Rarita-Schwinegr fermion and the pure Rarita-Schwinger gas in Sec. \ref{IV} and discuss the nonlinear quantum Hall effects for the Rarita-Schwinger gas in reality.  In the fourth section, we consider the small $G$ expansion and find that the Hamiltonians are hermitian while the corresponding Hall conductance is still  nonlinearly quantized as the Landau index varies. The last section is devoted to our conclusions and perspectives. Some unsolved problems are addressed. In the Appendix \ref{appen}, we rederive the quantum Hall conductance of spin-$\frac{1}{2}$ Dirac fermions by solving the Dirac equation under a constant electromagnetic field and perturbation theory. We use the same strategy to calculate the Hall conductance of the Rarita-Schwinger fermions. Its detailed calculations are presented in Appendix \ref{appen2}.

\section{Rarita-Schwinger fermions in a constant magnetic field} \label{III}

\subsection{Superluminal mode}

Before calculating the  Hall conductance of the Rarita-Schwinger gas, we first review the Rarita-Schwinger fermions coupled minimally to a constant magnetic field  in (2+1)-dimensions to illustrate the existence of the superluminal mode for later convenience. We denote  $\gamma^0=\sigma^z,\gamma^1=i\sigma^y,\gamma^2=-i\sigma^x$ with $\sigma^i$ being the Pauli matrices. The corresponding metric is $g_{\mu\nu}={\rm diag}(1,-1,-1)$. The gauge potential is chosen as $A^\mu=(Ex,0,-Bx)$, where $E$ and $B$ are constants. We set $\hbar=1$ and the electric charge $e=1$. The "speed of light"  $v_f$, which is the genuine speed of light in the vacuum in high energy physics while it is the Fermi velocity for a condensed matter system, is also set to be unity. 

 The Lagrangian for the Rarita-Schwinger fermions in (2+1)-dimensions can be written in the following form\cite{2drs},
\begin{equation}
L=-\bar{\psi}^\mu(\Gamma\cdot D+mB)_\mu^\nu\psi_\nu,\label{1v2}
\end{equation}
where $\bar{\psi}_\mu=\psi_\mu^\dagger\gamma^0$, $ D_\mu=i\partial_\mu+eA_\mu$ ; $\Gamma\cdot D$ is the shortness of $\Gamma^\mu D_\mu$; and the matrices $(\Gamma^\alpha)_\mu^\nu$ and $B_\mu^\nu$ are given by
\begin{eqnarray}
(\Gamma^\alpha)_\mu^\nu&=&\gamma^\alpha g_\mu^\nu+W(\gamma_\mu g^{\nu\alpha}+g_\mu^\alpha \gamma^\nu)-V\gamma_\mu \gamma^\alpha \gamma^\nu,\\
B_\mu^\nu&=&-(g_\mu^\nu-T\gamma_\mu \gamma^\nu)
\end{eqnarray}
with
\begin{equation}
V=-(3W^2+2W+1)/2,\quad T=3W^2+3W+1,
\end{equation}
where $W$ is an arbitrary number except for $W=1/2$. Here we choose $W=-1$, then, the Lagrangian (\ref{1v2}) becomes,
\begin{eqnarray}
L&=&-\bar{\psi}^\mu(g_\mu^\nu(\gamma\cdot D)-(\gamma_\mu D^\nu+ D_\mu\gamma^\nu)+\gamma_\mu(\gamma\cdot D)\gamma^\nu\nonumber\\
&&-m(g_\mu^\nu-\gamma_\mu\gamma^\nu))\psi_\nu.\label{5v2}
\end{eqnarray}
And the equation of motion {reads},
\begin{eqnarray}
L_\mu&\equiv&(g_\mu^\nu(\gamma\cdot D)-(\gamma_\mu D^\nu+ D_\mu\gamma^\nu)+\gamma_\mu(\gamma\cdot D)\gamma^\nu\nonumber\\
&&-m(g_\mu^\nu-\gamma_\mu\gamma^\nu))\psi_\nu=0.\label{6v2}
\end{eqnarray}
Note that the zeroth component of Eq. (\ref{6v2}) does not contain any time derivative, therefore it becomes a primary constraint. Contracting Eq. (\ref{6v2}) with $\gamma^\mu$ and $ D^\mu$ successively from the left hand side, one obtains 
\begin{eqnarray}
\gamma^\mu L_\mu
&=&(\gamma\cdot D)(\gamma\cdot\psi)-( D\cdot\psi)+2m(\gamma\cdot\psi)=0,\\\label{7v2}
 D^\mu L_\mu
&=&-ie\gamma\cdot F\cdot\psi+\frac{ie}{2}\gamma\cdot F\cdot\gamma(\gamma\cdot\psi)\nonumber\\
&&-2m^2(\gamma\cdot\psi)=0.\label{23}
\end{eqnarray}
Therefore we have the subsidiary condition,
\begin{equation}
\gamma\cdot\psi=\frac{ie}{2m^2}(-\gamma\cdot F\cdot \psi+\frac{1}{2}\gamma\cdot F\cdot \gamma(\gamma\cdot\psi)).\label{9v2}
\end{equation}
Substituting Eq. (\ref{9v2}) back into Eq. (\ref{6v2}), 
\begin{eqnarray}
L_\mu&=&(\gamma\cdot D-m)\psi_\mu-( D_\mu+m\gamma_\mu)(\gamma\cdot\psi)\nonumber\\
&=&(\centernot{ D}-m)\psi-( D_\mu+m\gamma_\mu)\frac{ie}{2m^2}(-\gamma\cdot F\cdot \psi\nonumber\\
&&+\frac{1}{2}\gamma\cdot F\cdot \gamma(\gamma\cdot\psi)).\label{10v2}
\end{eqnarray}
For $E=0$ and a constant magnetic field $F^{12}=B$, 
 Eq. (\ref{10v2}) becomes,
\begin{equation}
(\centernot{D} -m)\psi^\nu-G(m\gamma^\nu+D^\nu)\psi^0=0,\label{27}
\end{equation}
where $G=\frac{e B}{2m^2}$. {Equation (\ref{27}) tells us} that $\psi^0$ decouples from $\psi^1$ and $\psi^2$ fields while $\psi^1$ and $\psi^2$ couple together. Thus,  we consider the equation of motion for $\psi^0$ 
\begin{equation}
(i\gamma_0 \partial^0 +\gamma_i D^i-m)\psi^0-G(m\gamma^0+i\partial^0)\psi^0=0.\label{eom0}
\end{equation}
Defining $a=-\frac{D_1+ iD_2}{\sqrt{2eB}}$  which satisfies the canonical commutation relation $[a,a^\dagger]=1$,  Eq. (\ref{eom0}) becomes
\begin{equation}
i\partial_0\psi^0
=H_0\psi^0=m\left(
\begin{array}{ccc}
\frac{1+G}{1-G} & \frac{2\sqrt{G}a^\dagger}{1-G}  \\
\frac{2\sqrt{G}a}{1+G} & -\frac{1-G}{1+G}
\end{array}\right)\psi^0,\label{13}
\end{equation}
where $H_0$ then is the Hamiltonian of $\psi^0$. It is nonhermitian. The energy spectrum of $H_0\psi_0=E\psi_0$ is given by
\begin{equation}
E(\pm,n)=m\frac{2G\pm\sqrt{(1+G^2)^2+4G(1-G^2)n}}{|1-G^2|},\label{spectrum}
\end{equation}
where $+(-)$ stands for particles(holes). 

We see that the non-hermitness of (\ref{13}) does not affect the stability of the system when $G^2<1$ because the spectrum (\ref{spectrum}) is real.   $E(\pm,n)$ is divergent at $|G|=1$. This is in fact artificial and we will discuss this matter in the subsection C.

When $G^2>1$, the effect of the non-hermitness appears gradually. The square root in the numerator of (\ref{spectrum}) may be imaginary for a given $n$ and then this Landau level becomes unstable. While the lowest ($n=0$) and second ($n=1$) Landau levels are always stable, the higher Landau levels may be unstable. For example, when $1.3<G<7.6$, $E_2$ is complex; when $1.2<G<11.7$, $E_3$ is complex; and when $1.14<G<15.8$, 
$E_4$ is complex; and so on. Any $n>1$ Landau level may unstable in a given range of $G^2$. The larger $n$ is, the larger range of $G^2$ is. For $n\to\infty$, this range is $G^2\in (1,\infty)$. As we will see later in the end of this subsection, $\psi_0$ is superluminal in all spacelike intervals.

For later convenience, we will only consider the particle case in the following with the eigen wave functions
\begin{equation}
\psi^0(n):=
\left(
\begin{array}{ccc}
 \psi^0_\uparrow(n)  \\
 \psi^0_\downarrow(n-1)
\end{array}
\right)
=e^{-iE(n)t}
\left(
\begin{array}{ccc}
 \alpha^0_n\chi_n  \\
 \beta^0_{n-1}\chi _{n-1}
\end{array}
\right),\label{15}
\end{equation}
where $\chi_n$ is the wave function of the n-th Landau level, and $\alpha^0(\beta^0)$ is the upper(lower) coefficient of the spin component. If we denote $\chi_{-1}=0$, then Eq. (\ref{15}) can also describe the corresponding wave function for the case of $n=0$, the same situation as that of the Dirac fermion(see appendix \ref{appen}).

 Although {Lagrangian (\ref{5v2}) does not contain dynamics for $\psi^0$},  its dynamics is induced by the constraint (\ref{23}). On the other hand, $H_0$ depends on the choice of $W$ in the Lagrangian (\ref{1v2}), but the energy spectrum is independent of $W$. The expectation values of physical observables also do not depend on $W$.
 
 An analysis of the Green's function of the Rarita-Schwinger fermions shows that the propagator of $\psi^0$ may not vanish in a spacetime interval with \cite{2drs}
  \begin{eqnarray}
 \Delta r^2-\Delta t^2<G^2\Delta r^2. \label{spl}
 \end{eqnarray}
 This means that in the spacelike intervals with $0<\Delta r^2-\Delta t^2<G^2\Delta r^2$,  the $\psi_0$ is a superluminal mode if $G\ne 0$ ($B\ne 0$). There are still some spacelike intervals where $\Delta r^2-\Delta t^2>G^2\Delta r^2$ and $\psi_0$ does not propagate if $G^2<1$.  When $G^2>1$, Eq. (\ref{spl}) always holds.   The $\psi_0$ can propagate in all spacetime intervals either spacelike or timelike.

 \subsection{Total angular momentum representation}
 
Because of the constraints Eq. (\ref{9v2}) and $L_0$ of Eq. (\ref{6v2}), the $\psi^1$ and $\psi^2$ fields are connected with $\psi^0$ field. To demonstrate this more transparently, we shall rotate the $\psi^{1,2}$ fields into the total angular momentum representation\cite{ly,tlyw}, 
\begin{equation}
\left(
\begin{array}{cccc}
 \phi_{3\uparrow}  \\
 \phi_{3\downarrow}\\
 \phi_{1\uparrow}  \\
  \phi_{1\downarrow}
\end{array}
\right)
=\frac{1}{\sqrt{2}}
\left(
\begin{array}{cccc}
 -1 & 0 & i & 0  \\
 0  & 1 & 0 & i  \\
 0  & -1 & 0 & i  \\
 1 & 0 & i & 0  \\
\end{array}
\right)
\left(
\begin{array}{cccc}
 \psi_{1\uparrow}  \\
 \psi_{1\downarrow}\\
 \psi_{2\uparrow}  \\
  \psi_{2\downarrow}
\end{array}
\right),\label{trans}
\end{equation}
where $\phi_{3,\uparrow\downarrow}$ are the $J_z=\pm\frac{3}{2}$ components while $\phi_{1,\uparrow\downarrow}$ are the $J_z=\pm\frac{1}{2}$ components. Then, the constraints Eq. (\ref{9v2}) and $L_0$ can be rewritten as,
\begin{equation}
\gamma\cdot\psi=\gamma^0\psi_0-\sqrt{2}\phi_1=G\psi_0.\label{cons2}
\end{equation}
\begin{equation}
\gamma^0\phi_1=
\left(
\begin{array}{cccc}
 \sqrt{G}a & 0  \\
 0 & \sqrt{G}a^\dagger
\end{array}
\right)\phi_3+
\left(
\begin{array}{cccc}
 0 & \sqrt{G}a^\dagger  \\
 \sqrt{G}a & 0
\end{array}
\right)\phi_1.\label{cons1}
\end{equation}
The constraint (\ref{cons1}) tells us the relationship between $\phi_1$ and $\phi_3$, 
\begin{widetext}
\begin{equation}
\phi_1=\left(
\begin{array}{cccc}
 1+Ga^\dagger a & 0  \\
 0 & 1+Ga a^\dagger
\end{array}
\right)^{-1}\left(
\begin{array}{cccc}
 \sqrt{G}a  & -Ga^{\dagger 2}  \\
 -G a^2 & -\sqrt{G}a^\dagger
\end{array}
\right)\phi_3,\quad
\phi_3=\left(
\begin{array}{cccc}
 Ga^\dagger a & 0  \\
 0 & Ga a^\dagger
\end{array}
\right)^{-1}\left(
\begin{array}{cccc}
 \sqrt{G}a^\dagger  & -Ga^{\dagger 2}  \\
 -G a^2 & -\sqrt{G}a
\end{array}
\right)\phi_1.\label{R13}
\end{equation}
\end{widetext}
In terms of the constraints (\ref{cons1}) and (\ref{cons2}), $\phi_{1}$ and $\phi_3$ can be constructed from $\psi_0$ directly. From the Hamiltonian (\ref{13}) and the constraint (\ref{cons2}), the Hamiltonian $H_1$ for $\phi_1$ is given by,
\begin{equation}
H_1=m\left(
\begin{array}{ccc}
\frac{1+G}{1-G} & -\frac{2\sqrt{G}}{1+G}a^\dagger  \\
-\frac{2\sqrt{G}}{1-G}a & -\frac{1-G}{1+G}
\end{array}\right).\label{h1}
\end{equation}
The equations of motion of $\psi^{1,2}$ (\ref{27}) and the unitary transformation (\ref{trans}) give rise to the Hamiltonian $H_3$ for $\phi_3$,
\begin{equation}
H_3=m\gamma^0+\left[
\begin{array}{ccc}
\frac{2Gma^\dagger a}{(1-G)(1-G+Ga^\dagger a)} & \frac{-2Gm\sqrt{G}a^{\dagger 3}}{(1-G)(1-G+Ga^\dagger a)} \\
\frac{-2Gm\sqrt{G}a^{ 3}}{(1+G)(1+2G+Ga^\dagger a)} & -\frac{2Gma a^\dagger}{(1+G)(1+2G+Ga^\dagger a)}
\end{array}\right].\label{h3}
\end{equation}
According to the explicit construction of the Hamiltonian (\ref{h1}) and (\ref{h3}), one can verify that, the $\phi_1$ and $\phi_3$ states induced from the constrains (\ref{cons1}) and (\ref{cons2}) have the same energy as that of a given $\psi^0(n)$ state.  Although the constraints (\ref{cons1}) and (\ref{cons2}) produce the correct eigen functions for the Hamiltonian $H_1$ (\ref{h1}) and $H_3$ (\ref{h3}) from $\psi^0$ (\ref{15}) by  denoting $\chi_{-1}=\chi_{-2}=0$, the normalization is subtle. In order to have the correct normalization coefficients of all the fields, the conserved charge of the Rarita-Schwinger fermion is needed. The conserved current density of Lagrangian (\ref{5v2}) is given by
\begin{equation}
j^\nu=i\epsilon ^{\mu\nu\sigma}\bar{\psi}_\mu\psi_\sigma,\label{currentRS}
\end{equation}
and the conserved charge that is definitely positive (when $G<1$, $|\phi_3|>|\phi_1|$) is normalized to be unity
\begin{equation}
J^0=\langle \phi_3^\dagger \phi_3 -\phi_1^\dagger \phi_1 \rangle=1.\label{charge}
\end{equation}
Eq. (\ref{charge}) determines the normalizing condition for a single Rarita-Schwinger fermion, which is a (2+1)-dimensional reduction of the (3+1)-dimensional one \cite{Lu}.
We notice that this conserved current is of a different form from the familiar one 
\begin{eqnarray}
\tilde j^\mu=\bar\psi^\nu\gamma^\mu\psi_\nu.\label{cc}
\end{eqnarray}
In fact, it is easy to show that for a free Rarita-Schwinger gas, $j^\mu=\tilde j^\mu$ after the constraint $\gamma\cdot \psi=0$ is imposed. As the constraint alters after an external field is applied, these two currents are not equal and we must take the conserved one, eq. (\ref{currentRS}).

\subsection{Critical magnetic field}

As we have mentioned, the Rarita-Schwinger fermions have very different behaviors between $|G|>1$ and $|G|<1$. At the critical value $|G|=1$, in fact,  the equation of motion of $\psi^0$ becomes non-relativistic. To illustrate this point, we first rewrite Eq. (\ref{13}) as,
\begin{equation}
(I-G\gamma^0)i\partial_0\psi^0=(-\gamma^0\gamma^iD_i+m\gamma^0+Gm)\psi^0.\label{19}
\end{equation}
Equation (\ref{19}) can be interpreted as an equation of motion for a relativistic fermion with an anisotropic speed of light. When $G=1$, by denoting $\psi^0=(\alpha, \beta)^T$, then Eq. (\ref{19}) becomes,
\begin{equation}
\left\{
\left(
\begin{array}{ccc}
0 & 0  \\
0 & 2
\end{array}\right)i\partial^0
+
\left(
\begin{array}{ccc}
0 & D_-  \\
D_+ & 0
\end{array}\right)
-
\left(
\begin{array}{ccc}
2m & 0  \\
0 & 0
\end{array}\right)
\right\}
\left(
\begin{array}{ccc}
\alpha  \\
\beta
\end{array}\right)
=0,
\end{equation}
which means,
\begin{eqnarray}
2m\alpha-D_-\beta&=&0,\label{40v2}\\
D_+\alpha+2i\partial^0\beta&=&0.
\end{eqnarray}
These equations can be rewritten as,
\begin{equation}
i\partial_0\beta=H\beta=-\frac{D_i^2+eB}{2\tilde m}\beta,\label{42v2}
\end{equation}
which becomes a non-relativistic equation, with effective mass being $\tilde{m}=2m$. The energy is negative because Eq. (\ref{42v2}) represents the hole branch. When $G=-1$, the corresponding equation of motion describes the particle branch and its energy is positive.

The emergence of the non-relativistic Schr\"odinger equation (\ref{42v2}) from a relativistic Dirac equation (\ref{19}) lies in Eq. (\ref{40v2}). When $G=1$, the $\alpha$ component has no dynamics and Eq. (\ref{40v2}) reduces to a constraint equation. Therefore, only $\beta$ component is dynamic and becomes non-relativistic.

 By recovering all the physical constants, the critical magnetic strength is estimated by
\begin{equation}
G_c=\frac{v_f^2\hbar e B_c}{2m^2v_f^4}=1,\label{criticalB}
\end{equation}
where $v_f$ is the Fermi velocity and $mv_f^2$ is the band gap in materials. We take the graphene's data to estimate the critical field. The Fermi velocity in graphene is $10^6m/s\sim0.003c $\cite{gra1,gra2} and the band gap can be continuously tuned from $0eV$ up to $0.25eV$ in {a} bilayer graphene \cite{gra3}. Therefore, the critical magnetic strength $B_c$ in a graphene like material has a range form $0T$ to $10^2T$. Thus, the instability of higher Landau levels of the Rarita-Schwinger fermions is experimentally detectable.  On the other hand,  the fundamental Rarita-Schwinger fermion in the vacuum ($v_f=c$), though it has not been observed yet, has the critical $B\sim10^{10}T$ if it is assumed to be as light as an electron.

\section{Nonlinear Quantum Hall effects of Rarita-Schwinger gas}\label{IV}
Now we are ready to calculate the Hall conductance of the Rarita-Schwinger gas. In the first three subsections, we limit $|G|<1$.

\subsection{Single Rarita-Schwinger fermion's contribution to Hall conductance  }

Applying a small external constant electric field $E$ along the $x$-direction which is much smaller than the magnetic field, the constraint (\ref{9v2}) becomes,
\begin{equation}
\gamma\cdot\psi=G\psi^0+K\psi^2,\label{28}
\end{equation}
with $K\equiv \frac {eE}  {2m^2}$, while the other constraint (\ref{cons1}) remains the same. In the rotated representation (\ref{trans}), the constraint (\ref{28}) can be written as,
\begin{equation}
\gamma^0 \psi_0-\sqrt{2} \phi_1=G\psi_0-\frac{i}{\sqrt 2}K\phi_3+\frac{i}{\sqrt 2}K\gamma^2 \phi_1.\label{newcons1}
\end{equation}
{Substituting the constraint (\ref{28}) into (\ref{10v2}), the equation of motion of  $\psi^0$, the corresponding Hamiltonian $\tilde{H}_0$ becomes,
	\begin{equation}
		\tilde{H}_0=H_0+\tilde{H}_0',
	\end{equation}
	where $H_0$ is the unperturbed Hamiltonian (\ref{13}) {while the perturbed Hamiltonian $\tilde{H}_0'$ is given by},
	\begin{equation}
		\tilde{H}_0'=-E(1+O(E/B))x.
	\end{equation}
	Since  we consider the linear response to the external electric field,  the explicit form of $O(E/B)$ is not important. (For details, see Appendix \ref{appen2}.)

Similar to the Dirac fermion discussed in the Appendix \ref{appen}, we use the non-degenerate perturbation theory. The perturbed $\Psi^0$ is given by
\begin{eqnarray}
\Psi_0(n)&=&\psi_0(n)+\sum_m\frac{\langle n|(-Ex)|m\rangle}{E(n)-E(m)}\psi_0(m)\nonumber\\
&=&\psi_0(n)+\psi_0'(n+1)+\psi_0''(n-1),
\end{eqnarray}
where $\langle n|(-Ex)|m\rangle$ is the matrix element between $\psi_0^\dagger(n)$ and $\psi_0(m)$; $\psi_0'(n+1)$ and $\psi_0''(n-1)$ are the corresponding first order corrections to the unperturbed wave function $\psi_0(n)$ of the $n\pm1$ Landau levels. After calculating the perturbed wave function $\Psi_0$, we  obtain the perturbed $\Phi_1$ and $\Phi_3$ through the constraints (\ref{cons1}) and (\ref{cons2}). It is straightforward to calculate the Hall currents for a single Rarita-Schwinger fermion by substituting the perturbed $\Psi_0$, $\Phi_1$, and $\Phi_3$ into the expression for currents (\ref{currentRS}), namely,
\begin{eqnarray}
J^1&=&\langle i\bar{\psi}_0\psi_2-i\bar{\psi}_2\psi_0\rangle\nonumber\\
&=&\frac{1}{\sqrt{2}}\langle\Psi_{0\uparrow}^\dagger\Phi_{3\uparrow}+\Psi_{0\uparrow}^\dagger\Phi_{1\downarrow}-\Psi_{0\downarrow}^\dagger \Phi_{3\downarrow}-\Psi_{0\downarrow}^\dagger\Phi_{1\uparrow}\rangle\nonumber\\
&&+h.c.=0,\\
J^2&=&\langle i\bar{\psi}_1\psi_0-i\bar{\psi}_0\psi_1\rangle\nonumber\\
&=&\frac{i}{\sqrt{2}}\langle\Psi_{0\uparrow}^\dagger\Phi_{3\uparrow}-\Psi_{0\uparrow}^\dagger\Phi_{1\downarrow}+\Psi_{0\downarrow}^\dagger \Phi_{3\downarrow}-\Psi_{0\downarrow}^\dagger\Phi_{1\uparrow}\rangle\nonumber\\
&&+h.c.=-K(1+n)(4+12G+\mathcal{O}(G^2)),
\end{eqnarray}
where we have expanded to the first order of $G$ {(details are presented in Appendix \ref{appen2})}. We plot the Hall current of the Rarita-Schwinger fermion versus $G$ in Fig. \ref{J2G} and  K in Fig. \ref{J2K}. As a comparison, we also plot the  Hall current of the Dirac fermion for both cases. varying  the Hall current of the Rarita-Schwinger fermion as $G$ is dramatically different from that of the Dirac fermion: the former increases while the latter decreases as $G$ increases. The $K$-dependence of both Hall currents are similar as expected by the linear response to the electric field.

\begin{figure}
	\includegraphics[width=0.45\textwidth]{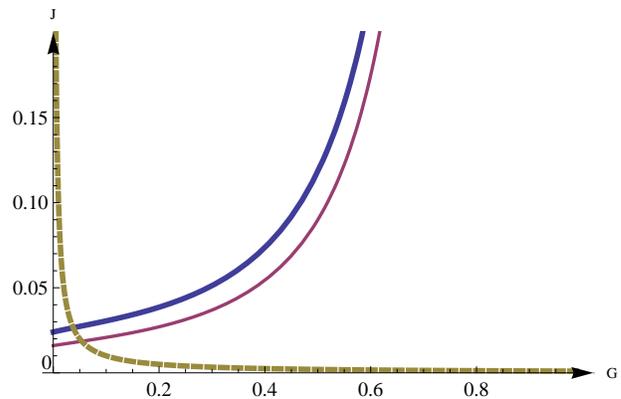}
	\caption{(color online)
	The blue(thick) and purple(solid) curves are the Hall currents versus $G$ for a single Rarita-Schwinger fermion. The brown(dashed) curve is that for a single Dirac fermion. The parameters are chosen as $n=5$ for blue and $n=3$ for purple while $K=0.001$. The perturbation theory fails for the Rarita-Schwinger fermion at large $G$ because the first order correction is larger than the unperturbed wave function. Therefore, the Hall current blows up at large $G$ and renders the plot unreliable.
	}
	\label{J2G}	
\end{figure}
\begin{figure}
	\includegraphics[width=0.45\textwidth]{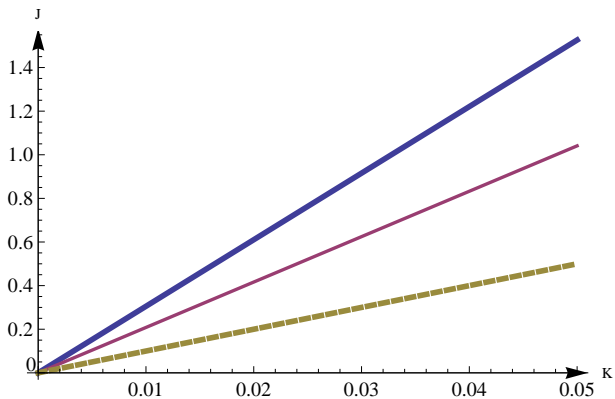}
	\caption{(color online)
		The blue(thick) and purple(solid) curves are the Hall currents versus $K$ for a single Rarita-Schwinger fermion. The brown (dashed) curve is that for a single Dirac fermion. The parameters are chosen as $n=5$ for blue and $n=3$ for purple while $G=0.1$.
	}
	\label{J2K}	
\end{figure}

Unlike the Hall current for a single Dirac fermion (\ref{halldirac1}) which is merely linearly dependent on the filling factor, we observe that the Hall current for a single Rarita-Schwinger fermion not only depends on the Landau level index $n$ and its mass $m$, but also contains a combination of $E\cdot B$ instead of $\frac{E}{B}$. The contribution to the Hall conductance from a single Rairata-Schwinger fermion is then given by 
\begin{equation}
\sigma_{xy}^{sp}=(\frac{e}{2m^2} (1+n)(4+12G+\mathcal{O}(G^2))+\frac{3}{2})\frac{e^2}{h},\label{hallRSs}
\end{equation}
where the last term  comes from the chiral anomaly of the Rarita-Schwinger fermion which is three times larger than that of the Dirac fermion\cite{GG,anomaly1,anomaly2}.

\subsection{Step-like but not plateaued Hall conductance of a pure Rarita-Schwinger gas}

Through the same argument leading to the  Hall conductance (\ref{14}) for the Dirac fermions in the Appendix \ref{appen}, the Hall conductance for the pure Rarita-Schwinger fermions is given by
\begin{equation}
\sigma_{xy}^{pure}=\rho\sigma_{xy}^{sp}=(A (1+n)(4+\frac{12A}\nu+\mathcal{O}((\frac{A}\nu)^2))+\frac{3}{2})\frac{e^2}{h},\label{hallRSp}
\end{equation}
where $\rho$ is the charge density, $\nu$ is the filling factor of the Landau level; $n$ the Landau level index and $A=\frac{\pi e\rho}{m^2}$. Since the system is not stable when $G\geq 1$, the filling factor $\nu$ is restricted to $\nu>A$.
Unlike the linear dependence on the filling factor for the non-relativistic fermions and Dirac fermions, the Hall conductance of the pure Rarita-Schwinger gas is exotic. For a given Landau level $n$, the Hall conductance decreases nonlinearly as the filling factor increases while there is a jump between two adjacent Landau levels (Fig. \ref{sigma}). We choose a graphene like parameters\cite{gra1,gra2} to plot Fig. \ref{sigma}, i.e., the Fermi velocity is chosen as $0.003c$, the mass gap is $0.06eV$ and the surface carrier density is of the order of $10^{11}$cm$^{-2}$. This gives $A\approx 0.1$. The stability of all Landau levels of  the system requires $G<1$ and then the filling factor $\nu>A\approx 0.1$. We would like to point out that the calculations nearby $\nu=0.1$ is not so reliable. Thus, for the lowest Landau level, we plot from $\nu=0.3$ in Fig. \ref{sigma}. 

\begin{figure}
	\includegraphics[width=0.45\textwidth]{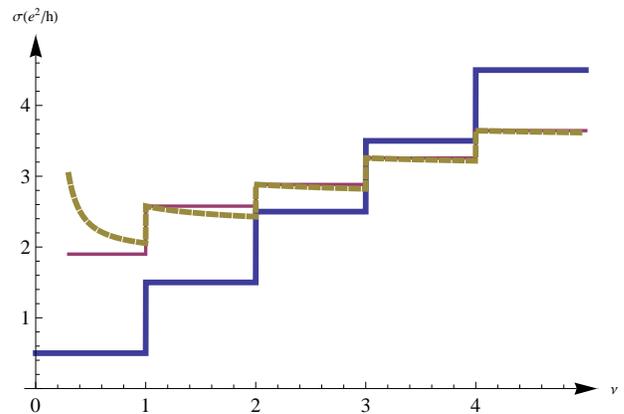}
	\caption{{(color online)
The nonlinear quantum Hall conductance of the Rarita-Schwinger gas. The purple (solid) curve is the quantum Hall conductance $\sigma$ (in unit of $\frac{e^2}{h}$) versus the filling factor $\nu$ for the Rarita-Schwinger fermion. The blue (thick) curve is the Hall plateaus for the Dirac fermion. The brown (dashed) curve is for the pure Rarita-Schwinger gas. The filling factor starts from $\nu=0.3$ as explained in the text. }}
	\label{sigma}	
\end{figure}

\subsection{Nonlinear dependence of the integer Landau index in the quantum Hall effects of the Rarita-Schwinger gas}

With these calculations for the single Rarita-Schwinger fermion and the pure Rarita-Schwinger gas in mind, we now consider a two-dimensional charged Rarita-Schwinger fermion gas in a condensed matter system. According to the structure of the wave function (\ref{15}), the degeneracy for each Landau level of the Rarita-Schwinger fermion is the same as that of the Dirac fermion (\ref{a11}).  For a conventional non-relativistic electron gas or a Dirac gas in two dimensions, the quantum states in the zero field limit are localized by the disorder. In a strong magnetic field, the extended states appear nearby the Landau level centers while the states away from the Landau level centers are localized. This leads to the quantum Hall effects in these systems.  The Hall conductance changes by an integer with the unit $\frac{e^2}h$ as the filling factor crosses two adjacent Landau levels. 

Although the Rarita-Schwinger gas under a disordered potential is a highly non-trivial problem which we are unable to solve it yet, we believe the general argument of the localization for a two-dimensional system can also be applied to the Rarita-Schwinger gas in the zero magnetic field. Due to the wave functions of the signle Rarita-Schwinger fermion basically consist of the conventional Landau level wave functions,  we may also assume the extended states appear nearby the Landau level center while the other states are still localized. We then can have a subversive quantum Hall conductance of the Rarita-Schwinger gas
\begin{equation}
\sigma_{xy}^{RSG}=(A(1+n)(4+\frac{12A}n+\mathcal{O}((\frac{A}n)^2))+\frac{3}{2})\frac{e^2}{h}\label{hallRS}
\end{equation} 
for $n\ne 0$. For the lowest Landau level ($n=0$), due to the system is not stable at $\nu\approx 0$, we are not able to know if there are the conventional extended states in-between $\nu=0$ and 1 under the disorder. We then can not fix the exact plateau position. If the extended states exist, we choose to neglect ${\cal O}(A^2)$ term in (\ref{hallRSp}), i.e., $\sigma_{xy}^{RSG}(n=0)=(4A+\frac{3}{2})\frac{e^2}{h}$ (see Fig. \ref{sigma}). Another possibililty is that there is no conventional extended states due to the localization. Then the Hall conductance in the lowest Landau level is merely given by the chiral anomaly. In sum,  we see the  quantum Hall effect for a given Landau level index in the sense that the Hall conductance is still a plateau while the magnitude of the plateaus is nonlinearly dependent on the integer Landau level filling factor as shown in Fig. \ref{sigma}. 

\subsection{Discussions}

Notice that the current (\ref{currentRS}) of the Rarita-Schwinger fermion implies that the electric currents $J^{1,2}$ are always related to the superluminal mode $\psi^0$,  the physical origin for these exotic Hall conductances (Eqs.(\ref{hallRSs}),fgas(\ref{hallRSp}) and (\ref{hallRS})) arises from the contribution of the superluminal mode. This result can also be expected through the existence of the critical magnetic strength (\ref{criticalB}) while the quantum Hall conductance does not depend on the strength of the external magnetic field {for Dirac fermions (\ref{a14}). 

We do not discuss the system with $|G|>1$ yet. In this case, the superluminal mode propagates in all spacelike intervals and the Landau levels are cut off at a finite integer. For example, if $11.7<G<15.8$,  only $n=0,1,2,3$ Landau levels are stable and the nonlinear plateaus with $n=0,1,2,3$ are left in Fig. \ref{sigma} although the exact values of the plateau positions need to be recalculated. (We do not do this here.)  If $7.6<G<11.7$, $n=0,1,2$ plateaus are left.  If $1.3<G<7.6$, $n=0,1$ plateaus are left. If $1.2<G<1.3$,   again only $n=0,1,2$ nonlinear plateaus are left. If $1.14<G<1.2$,  $n=0,1,2,3$ plateaus are left, and so on. This means that the filling factor of the system for a given $G>1$ can not exceed the upper bound. Otherwise, the system becomes unstable.

\section{Small G expansion}

In this section, we consider the small G expansion of the Hamiltonians (\ref{13}), (\ref{h1}), and (\ref{h3}), we will show that, to the first order of G, the Hamiltonians become hermitian, while the quantum Hall conductance remains step-like and nonlinear. This suggests that non-hermicity of the Hamiltonian is not the origin for the nonlinear quantum Hall conductance of the Rarita-Schwinger gas. 

When $G\ll1$, the Hamiltonian $H_0$ (\ref{13}) can be expanded to the first order of G, and the result is an hermitian Hamiltonian $\hat H_0$, i.e.,
\begin{equation}
\hat{H}_0= m\left(
\begin{array}{ccc}
1+2G & 2\sqrt{G}a^\dagger  \\
2\sqrt{G}a & -1+2G
\end{array}\right),\label{h0G}
\end{equation}
with the spectrum
\begin{eqnarray}
\hat E(\pm,n)&=& m(2G\pm\sqrt{1+4Gn})\nonumber\\
&\sim &m(2G\pm(1+2Gn))+\mathcal{O}(G^2). 
\end{eqnarray} 
One can easily check that this spectrum is identical to $E(\pm,n)$ (\ref{spectrum}) up to the first order of $G$. Using the constraint (\ref{cons2}), the expanded Hamiltonian $\hat H_1$ reads 
\begin{equation}
\hat{H}_1= m\left(
\begin{array}{ccc}
1+2G & -2\sqrt{G}a^\dagger  \\
-2\sqrt{G}a & -1+2G
\end{array}\right),\label{h1G}
\end{equation}
which is the same as the first order expansion of Hamiltonian (\ref{h1}). Expand the constraint (\ref{R13}) to the first order of $G$,
\begin{equation}
\phi_1=\left(
\begin{array}{cccc}
 \sqrt{G}a  & -Ga^{\dagger 2}  \\
 -G a^2 & -\sqrt{G}a^\dagger
\end{array}
\right)\phi_3.\label{R13ex1}
\end{equation}
Then the Hamiltonian $\hat H_3$ becomes
\begin{equation}
\hat H_3=m\left(
\begin{array}{ccc}
1+2Ga^\dagger a & -2G\sqrt{G}{a^\dagger}^3  \\
-2G\sqrt{G}a^3 & -1-2Gaa^\dagger
\end{array}\right),\label{h3G}
\end{equation}
with the spectrum $$\hat E_3(\pm,n)=2G\pm\sqrt{1+4Gn-G^3n+4G^2n^2+G^3n^3},$$ which is the same as $\hat E$ if we neglect the higher order terms of $G$ in the square root. As mentioned before, in the small $G$ region, all the Hamiltonians become hermitian. Similar to Sec. \ref{IV}, if we turn on a small electric field $E$, the perturbation Hamiltonian $H'=-Ex$ in the sense of linear response. Following the same treatment in the Appendix \ref{appen2}, the perturbed wave functions are given by
\begin{widetext}
\begin{eqnarray}
\hat \Psi_0&\sim& e^{-\hat E(n)t}\left(
\begin{array}{ccc}
\sqrt{G} \sqrt{2 (1 + n)}\chi_n-i \sqrt{2} K (1 + n) (2 + n)\chi_{n+1}-i K\sqrt{2n^3 (1 + n)}\chi_{n-1}  \\
G  \sqrt{2n (1 + n)}\chi_{n-1}-i \sqrt{2G}  K (1 + n)^{\frac{3}{2}} (2 + n)\chi_n-iK \sqrt{2Gn^3 ( n^2-1)}\chi_{n-2}
\end{array}\right),\\
\hat \Phi_1&\sim& e^{-\hat E(n)t}\left(
\begin{array}{ccc}
\sqrt{G (1 + n)}\chi_n  -i K (1 + n) (2 + n)\chi_{n+1} -i K\sqrt{n^3 (1 + n)}\chi_{n-1}  \\
-G \sqrt{n (1 + n)}\chi_{n-1} + i \sqrt G K (1 + n)^{\frac{3}{2}} (2 + n)\chi_n +iK \sqrt{Gn^3 ( n^2-1)}\chi_{n-2}
\end{array}\right),\\
\hat \Phi_3&\sim& e^{-\hat E(n)t}\left(
\begin{array}{ccc}
\chi_{n+1}-iK(n+1)\sqrt{\frac{n+2}{G}}\chi_{n+2}-iKn\sqrt{\frac{n+1}{G}}\chi_n \\
0
\end{array}\right), 
\end{eqnarray} 
\end{widetext}
where we have expanded to the leading order in each term, and they satisfy all the constraints and the normalization condition (\ref{charge})(to the leading order of G). Comparing with Eq. (\ref{b4}), (\ref{b5}), and (\ref{b6}), the wave functions are identical to the leading order. The corresponding single particle Hall currents are given by,
\begin{eqnarray}
J_1 &=&0,\\
J_2 &=&4K(1+n)+(20 + 38  n + 18 n^2) (G K).
\end{eqnarray}
Comparing with the Hall currents of the Dirac fermion (\ref{halldirac1}) we conclude that the quantum Hall conductance for the Rarita-Schwinger fermion is nonlinear even for an hermitian Hamiltonian. Since in the small $G$ region, the Rarita-Schwinger Hamiltonians (\ref{h0G}), (\ref{h1G}), and (\ref{h3G}) are hermitian, we expect that they are realizable in condensed matter systems\cite{tlyw}. 

\section{conclusions and perspectives}

In this paper, we calculated the Hall conductance of the massive spin-$\frac{3}{2}$ Rarita-Schwinger gas. 
We found that the result is not as simple as a combination of that of three copies of spin-$\frac{1}{2}$ Dirac fermions even for the single Rarita-Schwinger fermion. The pure Rarita-Schwinger gas presents step-like but non-plateaued Hall effects instead of the traditional Hall effects. In condensed matter systems, the Rarita-Schwinger gas also has quantum Hall effects but the Hall plateaus are not simply integer-valued but nonlinearly dependent on the Landau level index. For a strong enough magnetic field, the higher Landau levels may be unstable and then there is a upper bound of the filling factor beyond which the system is unstable.

To arrive at the quantum Hall effects, we have supposed that the Rarita-Scwinger gas is localized by the disorder in a zero field. This is in fact highly nontrivial and unsolved problem as the Rarita-Schwinger fermion in an external field may not be as easy as an electron gas or Dirac gas to be solved with a conventional way since the constraint on the Rarita-Schwinger fermion may be altered under the external potential.  The topological number for the Rarita-Schwinger gas was also not studied, so the topological stability of the quantum Hall plateau remains unclear. This is because defining the TKNN number \cite{TKNN} requires to introduce an external periodic potential which also needs to carefully deal with.  The edge states \cite{halp} were not studied because the edge potential is also an obstacle to solve the problem.

The authors are constructing a $k\cdot p$ Hamiltonian which contains the Rarita-Schwinger quasiparticles as the low-lying excitations \cite{tlyw}.  It was known for a long time that the properties of the Rarita-Schwinger fermions in an external potential are very interesting and sometimes puzzling \cite{rs1,2drs,rs2,weinberg,vz,potential,potential1,potential2}. We take a two-dimensional model as an example to  discuss the Klein paradox of the Ratira-Schwinger fermions in the tunneling through a step potential. We show that while the tunneling in the original Rarita-Schwinger theory must be helped by a novel soliton excitation which makes the wave function discontinuous at the step,  the soliton is replaced by the localized modes of  the $k\cdot p$ model at the step and the wave function is continuous. These localized states come from the spin-1/2 spinor which are not completely projected out.  We also propose possible real materials where the Rarita-Schwinger quasiparticles asymptotically emerge. The $k\cdot p$ theory may be generalized to a model in which the Rarita-Schwinger quasiparticles subjected to an external constant magnetic field and one expects that the quantum Hall conductance is nonlinear.  Finally, with the nonlinear quantum Hall conductance, we believe that there could be more (topological) structures or even a new paradigm (such as a new mechanism for non-Fermi liquids) hidden in higher spin systems, which are valuable for future study.

\acknowledgments
We thank Long Liang and Yong-Shi Wu for helpful discussions. This work was supported by  NNSF of China  11474061(XL,YY), NNSF of China 11525417 and 11374137 (FT,XGW).

\appendix

\section{Hall conductance of spin-$\frac{1}{2}$ Dirac fermions through perturbation theory}\label{appen}
In this appendix we shall give an example of calculating the Hall conductance of non-interacting Spin-$\frac{1}{2}$ Dirac fermions in (2+1)D\cite{mac}. Through this example, we will present the basic strategy of calculating the quantum Hall conductance for relativistic fermions which can be generalized to the case of the Rarita-Schwinger fermions. The Lagrangian for a free massive Spin-$\frac{1}{2}$ Dirac fermion moving under constant magnetic and electric field is given by,
\begin{equation}
L_{\frac{1}{2}}=\bar{\psi} (\gamma^\mu D_\mu-m)\psi,\label{1}
\end{equation}
where $D_\mu=i\partial_\mu+A_\mu$ with the equation of motion 
\begin{equation}
(\gamma^\mu D_\mu-m)\psi=0.
\end{equation}
The corresponding Hamiltonian
\begin{equation}
H=H_0+H'=-\gamma^0\gamma^1D_1-\gamma^0\gamma^2D_2+m\gamma^0-Ex,
\end{equation}
where we assume $H'=-Ex$ is a perturbation coming from a small enough external electric field $E$. The spectrum of the unperturbed Hamiltonian $H_0$ reads,
\begin{equation}
\varepsilon_0(n)=\pm\sqrt{2Bn+m^2},
\end{equation}
where $\pm$ reflects the particle-hole symmetry of the system. The unperturbed wave functions of the particle spectrum are given by,
\begin{equation}
\psi_0(n)=e^{-i\varepsilon_0(n)t}(\frac{m+\sqrt{m^2+2Bn}}{\sqrt{2Bn}}\phi(n),\phi(n-1))^T/N,\label{waveDirac}
\end{equation}
where $N$ is the normalizing factor and $\phi(n)$ is the wave function of the n-th Landau level. Though $n=0$ is a special case, the $n\rightarrow 0$ limit of the wave function (\ref{waveDirac}) gives the correct result by defining $\phi(-1)=0$.

There are many methods to determine the quantum Hall conductance for Dirac fermion\cite{mac}. Here we use the perturbation theory to calculate this which we also apply in calculating the quantum Hall conductance for the Rarita-Schwinger fermion. Define the creation and annihilation operator of the Landau levels,
\begin{equation}
a^\dagger \equiv (-D_1+iD_2)/\sqrt{2B},\quad a\equiv (-D_1-iD_2)/\sqrt{2B},
\end{equation}
then the perturbation $H'$ can be written as,
\begin{equation}
H'=i\frac{E}{\sqrt{2B}}(a^\dagger-a).
\end{equation}
From the standard non-degenerate perturbation theory, to the first order, the energy spectrum does not change while the wave function changes to,
\begin{equation}
\psi(n)=\psi_0(n)+\sum_m\frac{H'_{nm}}{\varepsilon_0(n)-\varepsilon_0(m)}\psi_0(m),\label{waveDirac2}
\end{equation}
where $H'_{nm}$ is the matrix element between the $\psi_0^\dagger(n)$ and the $\psi_0(m)$ states. The conserved current of the Dirac fermion is given by
\begin{equation}
J^\mu=\langle\bar{\psi}\gamma^\mu\psi\rangle.\label{currentDirac}
\end{equation}
Substituting the perturbed wave function (\ref{waveDirac2}) into the current (\ref{currentDirac}) and expanding to the leading order of $E/B$, the spatial components of the current becomes,
\begin{equation}
J^1(n)=0, \quad J^2(n)=-\frac{E}{B}. \label{halldirac1}
\end{equation}
 If the system has finite size with area $L_xL_y$, then the allowed minimal momentum jump is $\Delta p_y=2\pi/L_y$. Because the oscillating center $x_0=p_y/B$ of each Landau level wave function should satisfy $0<x_0<L_x$, we know that the degeneracy within each Landau level is given by,
\begin{equation}
\frac{BL_x}{\Delta p_y}=\frac{BL_x L_y}{2\pi}=\Phi/\Phi_0,\label{a11}
\end{equation}
which is the ratio between the total flux $\Phi$ and the flus quanta $\Phi_0$. If the filling factor of the system is some integral number $\nu$, then there are $\nu\Phi/\Phi_0$ particles in total which gives a surface density $\rho=\nu\Phi/\Phi_0/L_xL_y=\nu B/2\pi$. Therefore the Hall current $J_h$ reads
\begin{equation}
J_H(\frac{1}{2})=\rho J_2=\frac{\nu}{2\pi}E,
\end{equation}
and we get the  Hall conductance,
\begin{equation}
\sigma_{xy}(s=\frac{1}{2})=\frac{\nu}{2\pi}.\label{14}
\end{equation}
As a final remark of the appendix, we emphasize that the method presented in this section fails to capture the $\frac{1}{4\pi}$ contribution originated from the chiral anomaly of the zeroth Landau level\cite{GG}. Taking this fact into count, we will have the correct Hall conductance, namely,
\begin{equation}
\sigma_{xy}=(\nu+\frac{1}{2})\frac{e^2}{h}.\label{a14}
\end{equation}

\section{Detailed calculations on the quantum Hall conductance of Rarita-Schwinger fermion}\label{appen2}

The conserved current density derived from the Lagrangian (\ref{5v2}) of the Rarita-Schwinger fermion is given by
\begin{eqnarray}
j^\mu&=&-\bar{\psi}^\nu\gamma^\mu\psi_\nu+\bar\psi_\nu\gamma^\nu\psi^\mu+\bar\psi^\mu\gamma^\nu\psi_\nu-(\bar\psi\cdot\gamma)\gamma^\mu(\gamma\cdot\psi)\nonumber\\
&=&i\epsilon^{\nu\mu\rho}\bar{\psi}_\nu\psi_\rho, \label{b1}
\end{eqnarray}
where we have used the identity $\gamma^\mu \gamma^\nu=-i\epsilon^{\mu\nu\rho}\gamma_\rho+g^{\mu\nu}$. Using the transformation (\ref{trans}) between $\psi_{1,2}$ and $\phi_{1,3}$ fields, the conserved current density (\ref{b1}) can be rewritten as,
\begin{eqnarray}
j^0&=&\phi^\dagger_3\phi_3-\phi_1^\dagger\phi_1,\nonumber\\
j^1
&=&\frac{1}{\sqrt{2}}(\psi_{0\uparrow}^\dagger\phi_{3\uparrow}-\psi_{0\downarrow}^\dagger \phi_{3\downarrow}+\psi_{0\uparrow}^\dagger\phi_{1\downarrow}-\psi_{0\downarrow}^\dagger\phi_{1\uparrow})+h.c.,\nonumber\\
j^2
&=&\frac{i}{\sqrt{2}}(\psi_{0\uparrow}^\dagger\phi_{3\uparrow}+\psi_{0\downarrow}^\dagger \phi_{3\downarrow}-\psi_{0\uparrow}^\dagger\phi_{1\downarrow}-\psi_{0\downarrow}^\dagger\phi_{1\uparrow})+h.c.,\label{b2}\nonumber\\
\end{eqnarray}
where $\uparrow(\downarrow)$ stands for the upper(lower) component of the spinor field. Turning on the constant electromagnetic field, the constraint (\ref{9v2}) in the rotated representation (\ref{trans}) becomes,
\begin{equation}
\gamma^0 \psi_0-\sqrt{2} \phi_1=G\psi_0-\frac{i}{\sqrt 2}K\phi_3+\frac{i}{\sqrt 2}K\gamma^2 \phi_1,
\end{equation}
with $K\equiv \frac {eE}  {2m^2}$, while the other constraint (\ref{cons1}) does not change. Substitute the constraints back into the equation of motion of $\psi^0$ (\ref{10v2}), 
\begin{eqnarray}
&&i\partial_0(\gamma^0-G-(-\frac{i}{\sqrt{2}}K\square_1+\frac{i}{\sqrt 2}K\gamma^2)\frac{\gamma^0-G}{\square_2})\psi_0\nonumber\\
&=&(-Ex\gamma^0-D_i\gamma^i+m+GEx+Gm\gamma_0\nonumber\\
&&+(Ex+m\gamma_0)(-\frac{i}{\sqrt{2}}K\square_1+\frac{i}{\sqrt 2}K\gamma^2)\frac{\gamma^0-G}{\square_2})\psi_0,\label{A4}\nonumber\\
\end{eqnarray}
where
\begin{eqnarray}
\square_1&=&
\left(
\begin{array}{cccc}
 Ga^\dagger a & 0  \\
 0 & Ga a^\dagger
\end{array}
\right)^{-1}\left(
\begin{array}{cccc}
 \sqrt{G}a^\dagger  & -Ga^{\dagger 2}  \\
 -G a^2 & -\sqrt{G}a
\end{array}
\right)\\
\square_2&=&\sqrt 2-\frac{i}{\sqrt 2}K\square_1+\frac{i}{\sqrt{2}}K\gamma^2.
\end{eqnarray}
Let us estimate the contributions of each complicated terms in Eq. (\ref{A4}) to derive the linear response theory. Since $0<K\ll G<1$, especially when $G$ is small, the leading order in the operator $\square_1$ is proportional to $1/\sqrt G$ in the diagonal terms, and the leading order in $\square_2$ is $\sqrt 2$. Therefore to the leading order, the equation of motion (\ref{A4}) can be approximate as,
\begin{equation}
i\partial_0 \psi_0=(H_0+H_0')\psi_0,
\end{equation}
where $H_0$ is the unperturbed Hamiltonian (\ref{13}) and the perturbation $H_0'$ for $\psi^0$ field reads,
\begin{equation}
H_0'=-Ex=i\frac{E}{\sqrt{2B}}(a^\dagger-a),
\end{equation}
where $a=-\frac{D_1+iD_2}{\sqrt{2eB}}$ is the Landau level annihilation operator. For simplicity, we consider $|G|<1$ in the following. Using the non-degenerate perturbation theory, the perturbed wave function reads
\begin{eqnarray}
&&\Psi_0(n)=\psi_0(n)+\sum_m\frac{\langle n|H_0'|m\rangle}{E(n)-E(m)}\psi_0(m)\nonumber\\
&&=\left(
\begin{array}{ccc}
 \psi_{0\uparrow}(n)  \\
 \psi_{0\downarrow}(n-1)
\end{array}
\right)+
\left(
\begin{array}{ccc}
 \psi_{0\uparrow}'(n+1)  \\
 \psi_{0\downarrow}'(n)
\end{array}
\right)
+
\left(
\begin{array}{ccc}
 \psi_{0\uparrow}''(n-1)  \\
 \psi_{0\downarrow}''(n-2)
\end{array}
\right)\nonumber\\
&&\sim e^{-iE(n)t}(
\left(
\begin{array}{ccc}
(\sqrt{G} \sqrt{2 (1 + n)}+\mathcal{O}(G^{\frac 3 2}))\chi_n  \\
(G  \sqrt{2n (1 + n)}+\mathcal{O}(G^2))\chi_{n-1}
\end{array}
\right)\nonumber\\
&&+
\left(
\begin{array}{ccc}
(-i \sqrt{2} K (1 + n) (2 + n)+\mathcal{O}(K))\chi_{n+1}  \\
(-i \sqrt{2G}  K (1 + n)^{\frac{3}{2}} (2 + n)+\mathcal{O}(KG^{\frac 3 2}))\chi_n
\end{array}
\right)\nonumber\\
&&+
\left(
\begin{array}{ccc}
(-i K\sqrt{2n^3 (1 + n)}+\mathcal{O}(K))\chi_{n-1}  \\
(-iK \sqrt{2Gn^3 ( n^2-1)}+\mathcal{O}(KG^{\frac 3 2}))\chi_{n-2}
\end{array}
\right)),\label{b4}
\end{eqnarray}
where $\psi_0(n)$ is the eigen wave function (\ref{15}) of $H_0$ (\ref{13}) for a given Landau level $n$; $\psi'(n+1)$ and $\psi"(n-1)$ are the first order corrections to $\psi(n)$ where $n+1$ and $n-1$ means the corrections coming from the $n\pm1$ Landau levels. The last three lines in Eq. (\ref{b4}) are the leading order expansion of the wave function. Since we only consider the linear response of the external electric field, therefore through the constraints (\ref{newcons1}) and (\ref{cons1}),
\begin{eqnarray}
&&\Phi_1(n)=\frac{1}{\sqrt 2}(\gamma^0-G)\Psi_0(n)\nonumber\\
&&=\left(
\begin{array}{ccc}
 \phi_{1\uparrow}(n)  \\
 \phi_{1\downarrow}(n-1)
\end{array}
\right)+
\left(
\begin{array}{ccc}
 \phi_{1\uparrow}'(n+1)  \\
 \phi_{1\downarrow}'(n)
\end{array}
\right)
+
\left(
\begin{array}{ccc}
 \phi_{1\uparrow}''(n-1)  \\
 \phi_{1\downarrow}''(n-2)
\end{array}
\right)\nonumber\\
&&\sim e^{-iE(n)t}(
\left(
\begin{array}{ccc}
(\sqrt{G (1 + n)}+\mathcal{O}(G^{\frac 3 2}))\chi_n  \\
(-G \sqrt{n (1 + n)}+\mathcal{O}(G^2))\chi_{n-1}
\end{array}
\right)\nonumber\\
&&+
\left(
\begin{array}{ccc}
( -i K (1 + n) (2 + n)\chi_{n+1}+\mathcal{O}(K))\chi_{n+1}  \\
(  i \sqrt G K (1 + n)^{\frac{3}{2}} (2 + n)+\mathcal{O}(KG^{\frac 3 2}))\chi_n
\end{array}
\right)\nonumber\\
&&+
\left(
\begin{array}{ccc}
( -i K\sqrt{n^3 (1 + n)}+\mathcal{O}(K))\chi_{n-1}  \\
(iK \sqrt{Gn^3 ( n^2-1)}+\mathcal{O}(KG^{\frac 3 2}))\chi_{n-2}
\end{array}
\right)),\label{b5}
\end{eqnarray} 
where $\phi_1(n)$ is the unperturbed wave function. The perturbed $\Phi_3$ is related to $\Phi_1$ through Eq. (\ref{R13}), i.e.,
\begin{eqnarray}
 &&\Phi_3(n)=\left(
 \begin{array}{cccc}
  Ga^\dagger a & 0  \\
  0 & Ga a^\dagger
 \end{array}
 \right)^{-1}\left(
 \begin{array}{cccc}
  \sqrt{G}a^\dagger  & -Ga^{\dagger 2}  \\
  -G a^2 & -\sqrt{G}a
 \end{array}
 \right)\Phi_1(n)\nonumber\\
 &&=
 \left(
 \begin{array}{ccc}
  \phi_{3\uparrow}(n+1)  \\
  \phi_{3\downarrow}(n-2)
 \end{array}
 \right)+
 \left(
 \begin{array}{ccc}
  \phi_{3\uparrow}'(n+2)  \\
  \phi_{3\downarrow}'(n-1)
 \end{array}
 \right)
 +
 \left(
 \begin{array}{ccc}
  \phi_{3\uparrow}''(n)  \\
  \phi_{3\downarrow}''(n-3)
 \end{array}
 \right)\nonumber\\
&&\sim e^{-iE(n)t}(
\left(
\begin{array}{ccc}
(1+\mathcal{O}(1))\chi_{n+1}  \\
(0+\mathcal{O}(G^{\frac 3 2}))\chi_{n-2}
\end{array}
\right)\nonumber\\
&&+
\left(
\begin{array}{ccc}
(-iK(n+1)\sqrt{\frac{n+2}{G}}+\mathcal{O}(K)\chi_{n+2}  \\
(0+\mathcal{O}(K)\chi_{n-1}
\end{array}
\right)\nonumber\\
&&+
\left(
\begin{array}{ccc}
(-iKn\sqrt{\frac{n+1}{G}}+\mathcal{O}(K))\chi_{n}  \\
0+\mathcal{O}(K)\chi_{n-3}
\end{array}
\right)),\label{b6}
\end{eqnarray}
where $\phi_3(n)$ is the unperturbed wave function. Substituting the perturbed wave functions (\ref{b4}), (\ref{b5}), and (\ref{b6}) back into the current density (\ref{b2}), we have,
\begin{eqnarray}
j^1(n)
&=&\frac{1}{\sqrt{2}}({\psi'}_{0\uparrow}^\dagger(n+1)\phi_{3\uparrow}(n+1)+\psi_{0\uparrow}^\dagger(n){\phi''}_{3\uparrow}(n)\nonumber\\
&-&\psi_{0\downarrow}^\dagger(n-1) {\phi'}_{3\downarrow}(n-1)-{\psi''}_{0\downarrow}^\dagger(n-2) \phi_{3\downarrow}(n-2)\nonumber\\
&+&{\psi}_{0\uparrow}^\dagger(n){\phi'}_{1\downarrow}(n)+{\psi''}_{0\uparrow}^\dagger(n-1)\phi_{1\downarrow}(n-1)\nonumber\\
&-&{\psi}_{0\downarrow}^\dagger(n-1){\phi''}_{1\uparrow}(n-1)-{\psi'}_{0\downarrow}^\dagger(n)\phi_{1\uparrow}(n))+h.c.,\label{b7}\nonumber\\
\end{eqnarray}
\begin{eqnarray}
j^2(n)
&=&\frac{i}{\sqrt{2}}({\psi'}_{0\uparrow}^\dagger(n+1)\phi_{3\uparrow}(n+1)+\psi_{0\uparrow}^\dagger(n){\phi''}_{3\uparrow}(n)\nonumber\\
&+&\psi_{0\downarrow}^\dagger(n-1) {\phi'}_{3\downarrow}(n-1)+{\psi''}_{0\downarrow}^\dagger(n-2) \phi_{3\downarrow}(n-2)\nonumber\\
&-&{\psi}_{0\uparrow}^\dagger(n){\phi'}_{1\downarrow}(n)-{\psi''}_{0\uparrow}^\dagger(n-1)\phi_{1\downarrow}(n-1)\nonumber\\
&-&{\psi}_{0\downarrow}^\dagger(n-1){\phi''}_{1\uparrow}(n-1)-{\psi'}_{0\downarrow}^\dagger(n)\phi_{1\uparrow}(n))+h.c..\label{b8}\nonumber\\
\end{eqnarray}
Equations (\ref{b7}) and (\ref{b8}) directly tell us if the external electric field is zero, then $J^1=J^2=0$. Integrating over the entire space, the expectation values of the conserved current are given by,
\begin{eqnarray}
J_1=\langle j_1\rangle&=&0,\\
J_2=\langle j_2\rangle&=&K(1+n)(4+12G+\mathcal{O}(G^2)),
\end{eqnarray}  
where we have expanded $\langle J_2\rangle$ to the first order of $G$. From the structure of $\phi^0$ (\ref{15}), we know that if the energy of $\phi^0$ is real and forms Landau levels, then it will have the same Landau level degeneracy as that of the Dirac fermions. Therefore the surface density of the Rarita-Schwinger fermion is also given by
\begin{equation}
\rho=\nu B/2\pi,
\end{equation}
where $\nu$ is the filling factor. Then the Hall current reads,
\begin{equation}
J_H(\frac 3 2)=\rho J_2=\frac{\nu G (1+n)(4+12G+\mathcal{O}(G^2))}{2\pi}E.
\end{equation}
Therefore the corresponding Hall conductance for the Rarita-Schwinger fermion is given by,
\begin{equation}
\sigma_{xy}^{p}=(\nu G (1+n)(4+12G+\mathcal{O}(G^2))+\frac{3}{2})\frac{e^2}{h},\label{b12}
\end{equation}
where $\frac{3}{2}\frac{e^2}{h}$ comes from the chiral anomaly of the Rarita-Schwinger fermion\cite{GG,anomaly1,anomaly2}.


\begin{thebibliography}{99}


\bibitem{iqh} K. von Klitzing, G. Dorda, and M. Pepper, Phys. Rev. Lett. {\bf 45}, 494 (1980).
\bibitem{fqh} D. C. Tsui, H. L. Stormer, and A. C. Gossard, Phys. Rev. Lett. {\bf 48}, 1559 (1982).

\bibitem{km} C.L. Kane and E.J. Mele, Phys. Rev. Lett. {\bf 95}, 146802 (2005).

\bibitem{km1} C.L. Kane and E.J. Mele, Phys. Rev. Lett. {\bf 95},  226801(2005).


\bibitem{hald} F. D. M. Haldane, Phys. Rev. Lett. {\bf 61}, 2015(1988). 

\bibitem{qsh} M. Konig, S. Wiedmann, C. Brne, A. Roth, H. Buhmann, L. W. Molenkamp, X. L. Qi and S. C. Zhang, Science
{\bf 318}, 766 (2007).

\bibitem{qah}  C.-Z. Chang, J. Zhang, X. Feng, J. Shen, Z.  Zhang, M. Guo,  K. Li, Y. B. Ou, P. Wei, L.-L. Wang, Z.-Q. Ji, Y. Feng, S. H.  Ji, X. Chen, J. F.  Jia, X. Dai, Z. Fang, S.-C. Zhang,
K. He, Y. Y.  Wang, L. Lu, X.-C. Ma, and Q.-K. Xue,  Science {\bf 340}, 167 (2013).

\bibitem{wan} X.-G. Wan, A.M. Turner, A. Vishwanath, and S. Y. Savrasov, Phys.
Rev. B {\bf 83}, 205101 (2011). 

\bibitem{kz}  M. Z. Hasan and C. L. Kane,  Rev. Mod. Phys. {\bf 82}, 3045(2010).
\bibitem{qz} X. L. Qi and S. C. Zhang, Rev. Mod. Phys. {\bf 83}, 1057 (2011).


\bibitem{cglw} X. Chen, Z.-C. Gu, Z.-X. Liu, and X.-G. Wen, Phys. Rev. B {\bf 87}, 155114 (2013).

\bibitem{ki} A. Y. Kitaev, Annals of Physics
{\bf{303}}, 2 (2003).

\bibitem{lw} M. A. Levin and X. G. Wen, Phys. Rev. B {\bf 71}, 045110 (2005).




\bibitem{ly} L. Liang and Y. Yu, Phys. Rev. B {\bf 93}, 045113 (2016).

\bibitem{IF} H. Isobe, and L. Fu, Phys. Rev. B {\bf 93}, 241113 (2016).

\bibitem{newfe}  B. Bradlyn, J. Cano, Z. J.  Wang, M. G. Vergniory, C. Felser, R. J. Cava and B. A. Bernevig, Science {\bf 353}, 558 (2016).

\bibitem{hlf} T. H. Hsieh, J. Liu and L. Fu, Phys. Rev. B {\bf 90}, 08112(R) (2014).

\bibitem{wfdf1}  H. M.  Weng, C. Fang, Z. Fang, and X. Dai, Phys. Rev. B {\bf 93}, 241202 (2016).

\bibitem{wfdf2} H. M.  Weng, C. Fang, Z. Fang, and X. Dai, arXiv:1605.05186. 

\bibitem{zz} Z. M.  Zhu, G.  W. Winkler, Q. S. Wu, J. Li, and  A.  A. Soluyanov, Phys. Rev. X {\bf 6}, 031003 (2016).

\bibitem{hs} G. Q. Chang, S.- Y.  Xu, S.- M. Huang, D. S. Sanchez, C.-H. Hsu, G. Bian, Z.- M. Yu, I. Belopolski,
 N. Alidoust, H. Zheng, T.- R. Chang, H.- T. Jeng, S. Y.  A. Yang, T.  Neupert, H. Lin, and  M. Z.  Hasan, 
 arXiv:1605.06831.
 
 \bibitem{wkrk} B. J. Wieder, Y. Kim, A.M. Rappe, and C.L. Kane, Phys. Rev.
Lett. {\bf 116}, 186402 (2016). 
 
 \bibitem{ez} M. Ezawa, arXiv: 1609.03121.
 
\bibitem{tlyw} F. Tang, X. Luo, Y. Yu, and X.-G. Wan, to appear.

\bibitem{note} Dirac-Weyl type fermions with arbitrary spin in two-dimensional optical superlattices have been studied \cite{colda}. 

\bibitem{colda} Z. Lan, N. Goldman, A. Bermudez, W. Lu, and P.  \"Ohberg, Phys. Rev. B {\bf 84}, 165115 (2011) and references therein.

\bibitem{rs1} G. Velo and D. Zwanziger, Phys. Rev. {\bf 186} 1337 (1969).
\bibitem{2drs} M. Horta\ifmmode \mbox{\c{c}}\else \c{c}\fi{}su, Phys. Rev. D  {\bf 9}, 928 (1974).

\bibitem{rs2} M. Seetharaman, J. Prabhakaran, and P. M. Mathews, Phys. Rev. D {\bf 12}, 458 (1975).


\bibitem{vz} G. Velo and D. Zwanziger, Phys. Rev. {\bf 188}, 2218 (1969). 

\bibitem{potential} C. R. Hagen, Phys. Rev. D {\bf 4}, 2204 (1971).

\bibitem{potential1} L. P. S. Singh,  Phys. Rev. D {\bf 7}, 1256 (1973).

\bibitem{potential2} C. R. Hagen and L. P. S. Singh, Phys. Rev. D {\bf 26}, 393 (1982).

\bibitem{weinberg} See e.g., S. Weinberg, \textit{The Quantum Theory of Fields, Volume 1} (Cambridge University Press, 2005).





 


\bibitem{gra1} K. S. Novoselov, A. K. Geim, S. V. Morozov, D. Jiang, M. I. Katsnelson, I. V. Grigorieva, S. V. Dubonos and A. A. Firsov, Nature {\bf 438}, 197 (2005).

\bibitem{gra2} Y. B. Zhang, Y. -W. Tan, H. L. Stormer and P. Kim, Nature {\bf 438}, 201 (2005).

\bibitem{gra3} Y. B. Zhang, T. -T. Tang, C. Girit, Z. Hao, M. C. Martin, A. Zettl, M. F. Crommie, Y. R. Shen and F. Wang, Nature {\bf 459}, 820 (2009). 



 



\bibitem{Lu} D. Lurie, \textit{Particles and Fields} (John Wiley \& Sons Inc, 1968).




\bibitem{GG} V. P. Gusynin and S. G. Sharapov, Phys. Rev. Lett. {\bf 95}, 146801 (2005).

\bibitem{anomaly1} N. K. Nielsen, M. T. Grisaru, H. R\"{o}mer and P. Van Nieuwenhuizen, Nucl. Phys. B {\bf140}, 477 (1978).


\bibitem{anomaly2} L. Alvarez-Gaumi\'{e} and E. Witten,  Nucl. Phys.  B {\bf 234}, 269 (1983).


\bibitem{be1} Y. Hatsugai, Phys. Rev. Lett. {\bf 71}, 3697 (1993).

\bibitem{be2} S. Ryu and Y. Hatsugai, Phys. Rev. Lett. {\bf 89}, 077002 (2002).


\bibitem{be3} R. S. K. Mong and V. Shivamoggi, Phys. Rev. B {\bf 83}, 125109 (2011).

\bibitem{be4} J. Cano, M. Cheng, M. Mulligan, C. Nayak, E. Plamadeala, and J. Yard, Phys. Rev. B {\bf 89}, 115116 (2014).

\bibitem{TKNN}  Thouless D. J., M. Kohmoto, M. P. Nightingale, and M. den Nijs
, Phys. Rev. Lett. {\bf 49}, 405 (1982).
\bibitem{halp} B. I. Halperin, Phys. Rev. B {\bf 25}, 2185 (1982).

\bibitem{mac} A. H. MacDonald, Phys. Rev. B {\bf 28}, 2235 (1983).
\end{thebibliography}
\end{document}